\documentclass[10pt,letterpaper]{article}
\usepackage[top=0.85in,left=1in,footskip=0.75in]{geometry}

\usepackage{amsmath,amssymb}

\usepackage{changepage}

\usepackage{textcomp}

\usepackage{hyperref}

\usepackage[right]{lineno}

\usepackage[nopatch=eqnum]{microtype}
\DisableLigatures[f]{encoding = *, family = * }

\usepackage{array}

\usepackage[aboveskip=10pt,labelfont=bf,labelsep=period,justification=raggedright,singlelinecheck=off]{caption}

\usepackage[numbers,sort&compress]{natbib}
\usepackage{graphicx}
\usepackage{rotating}
\usepackage{subcaption}
\usepackage{amsthm}
\usepackage{url}
\usepackage{enumitem}
\usepackage{booktabs}
\usepackage{multirow}

\begin{document}
\vspace*{0.2in}

\begin{flushleft}
{\Large
\textbf\newline{Interpretable Role-Based Clustering in Multi-Layer Financial Networks}
}
\newline
\\
Christian Franssen\textsuperscript{1,*},
Thao Le\textsuperscript{1},
Iman van Lelyveld\textsuperscript{1,2},
Bernd Heidergott\textsuperscript{1}
\\
\bigskip
\textbf{1} VU Amsterdam
\\
\textbf{2} De Nederlandsche Bank \\
\bigskip

\noindent * Corresponding author

\end{flushleft}
\section*{Abstract}
Understanding the functional roles of financial institutions within interconnected markets is critical for effective supervision, systemic risk assessment, and resolution planning. We propose an interpretable role-based clustering approach for multi-layer financial networks, designed to identify the functional positions of institutions across different market segments. Our method follows a general clustering framework defined by proximity measures, cluster evaluation criteria, and algorithm selection. We construct explainable node embeddings based on egonet features that capture both direct and indirect trading relationships within and across market layers. Using transaction-level data from the ECB's Money Market Statistical Reporting (MMSR), we demonstrate how the approach uncovers heterogeneous institutional roles such as market intermediaries, cross-segment connectors, and peripheral lenders or borrowers. The results highlight the flexibility and practical value of role-based clustering in analyzing financial networks and understanding institutional behavior in complex market structures.

\noindent\textbf{Keywords:} financial networks; multi-layer networks; role-based clustering; node embeddings; interbank markets; systemic risk

\section{Introduction}

Understanding the structure of financial networks is essential for central banks seeking to monitor market functioning, assess systemic risk, and ensure the effective transmission of monetary policy. In these networks, financial institutions are modeled as nodes and transactions as directed edges, enabling detailed analysis of liquidity flows, intermediation chains, and cross-market dynamics \citep{bargigli2015multiplex, montagna2016multi}. Over the past two decades, financial network analysis has become a foundational tool in macroprudential supervision, with studies showing how the architecture of interbank markets influences contagion (e.g., see \citet{allen2000financial, battiston2012debtrank, staum2012counterparty, acemoglu2015systemic}), market segmentation \citep{ballensiefen2023money, 
EISENSCHMIDT2024103738}, and the effectiveness of central bank interventions \citep{EISENSCHMIDT2024103738, grasso2025flexible}. 

An important contribution to the study of money market networks is provided by \citet{CRAIG2014322}, who demonstrate that the German money market exhibits a core-periphery structure, with central institutions acting as intermediaries for those on the periphery. Similar structural patterns have been observed in other interbank markets, including those of the Netherlands and Brazil \citep{van2014finding, silva2016network}. \citet{carreno2017identifying} also postulate a core-periphery model on Chilean inter-bank exposures and find a dynamic main core, occasionally a secondary core, and several functionally distinct (net lending and borrowing) peripheries whose membership shifts over time and can differ across instruments. Moreover, \citet{Kojaku_2018} model the Italian e-MID (Electronic Market for Interbank Deposits) while allowing for multiple core–periphery structures and shows that the Italian e-MID network comprises several of these.

Understanding the roles that financial institutions occupy within such networks is crucial for assessing market functioning and regulatory oversight. For example, \citet{Kojaku_2018} find that the e-MID overnight market switched from multiple core–periphery pairs to a bipartite, broker-driven structure during the 2007-2009 financial crisis, re-assigning formerly peripheral banks into bridging positions. Such functional transitions can serve as indicators of market stress.
Moreover, role identification also shapes resolution policy: \citet{jackson2024credit} find that,  counter-intuitively, rescuing peripheral banks can be significantly more cost-effective than bailing out core institutions.
Relatedly, connectivity and substitutability are important considerations in determining whether a bank is classified as a globally systemic bank (G-SIB). Notably, G-SIB status comes with significantly higher compliance costs, thus stressing the importance of accurately identifying systemic roles within financial networks.

This research is motivated by the fact that over the past years, the structure of funding markets has become increasingly fragmented across instruments and segments  \citep{ecb2010, ecb2014, ecb2018, ecb2022}, most notably the unsecured and secured (repo) markets.
Institutions may engage in different capacities across these segments, acting as liquidity providers in one while borrowing in another, creating complex cross-market intermediation patterns that cannot be fully captured by single-layer analysis. While a single-layer analysis that merges both segments into one network can still identify highly connected intermediaries, it loses segment-level information by construction, so it is no longer possible to distinguish a bank's activity within each segment.
This underscores the importance of classifying institutional roles based on a joint analysis of multiple market segments, rather than assigning roles separately within each market. The traditional approach, categorizing institutions as either core or peripheral in isolation, overlooks the interconnected nature of their activities across markets.

In this paper, we contribute to the broader objective of identifying functional roles for financial institutions through an analysis of their positions across money market segments, allowing us to uncover cross-market intermediation structures that emerge only when the markets are studied together. 
The approach used in this paper is best understood as \textit{model-free} in the sense that it does not require a prespecified structure (core-periphery or otherwise). Instead, we use explainable node embeddings based on egonet features, i.e., localized substructures around each institution, allowing for the clustering of financial institutions by their functional position in the network. The result is that natural roles emerge via an interpretation of the embedding of clustered nodes.

We illustrate our approach using the ECB's Money Market Statistical Reporting (MMSR) dataset, which provides transaction-level data on secured and unsecured money market activity among euro-area institutions.
Focusing on a network of largest banks in each segment, we construct a clustering of institutions into six distinct groups, characterized by: (i) connectivity (systemic importance); (ii) funding balance (lending vs. borrowing); (iii) segment balance (segment specialization vs. cross-segment activity); and (iv) counterparty access (direct vs. indirect counterparty reach).
This clustering reveals institutions that act as intermediaries within or across markets, even if they are not highly connected in any single layer. The clear interpretability of these clusters in terms of functional roles makes the results directly useful for central banks and supervisors in understanding how liquidity and risk can transmit both within and across different segments of the money market.

The remainder of the paper is structured as follows. Section~\ref{sec:clust} introduces a general framework for graph clustering, allowing for well-known connectivity-based clustering of nodes in communities, and role-based clustering as used in this paper. Through a detailed step-by-step approach, we explain how clustering can be performed by: (1) identifying what defines proximity between nodes within graphs, (2) specifying the objectives for grouping proximate nodes into an optimal number of clusters, and (3) discussing various algorithms available to optimize these objectives. Section~\ref{sec:rol_mul_fin} outlines our methodology for interpretable role-based clustering in multi-layer financial networks, while also presenting a numerical example using MMSR data, and Section~\ref{sec:conc}  concludes with a discussion.

\section{Graph Clustering Pipeline} \label{sec:clust}

Clustering graphs is a broad and versatile field, with a range of approaches depending on how one defines a meaningful grouping of nodes. The framework presented here provides a clear and interpretable way to identify what constitutes effective graph clustering. In our view, setting up a cluster analysis is comprised of three main steps:
\begin{enumerate}[label=Step~\arabic*., leftmargin=*]
\item \textbf{Specify Node Proximity.}\
Define how to measure proximity between nodes, using either the graph's topological properties (e.g., shortest‐path or random‐walk distances), or node features (Step~1a) and construct a similarity metric (Step~1b), see Section~\ref{sec:prox}.

\item \textbf{Choose a Clustering Evaluation Metric.}\
Select an evaluation metric, aiming to maximize within, between, or both within and between cluster proximity (see Section~\ref{sec:ob}).

\item \textbf{Optimize with an Algorithm.}\
Choose a suitable algorithm to optimize the selected clustering evaluation metric in Step~2, see Section~\ref{sec:alg}.
\end{enumerate}

\begin{figure}[htbp]
    \centering
    \includegraphics[width=0.85\linewidth]{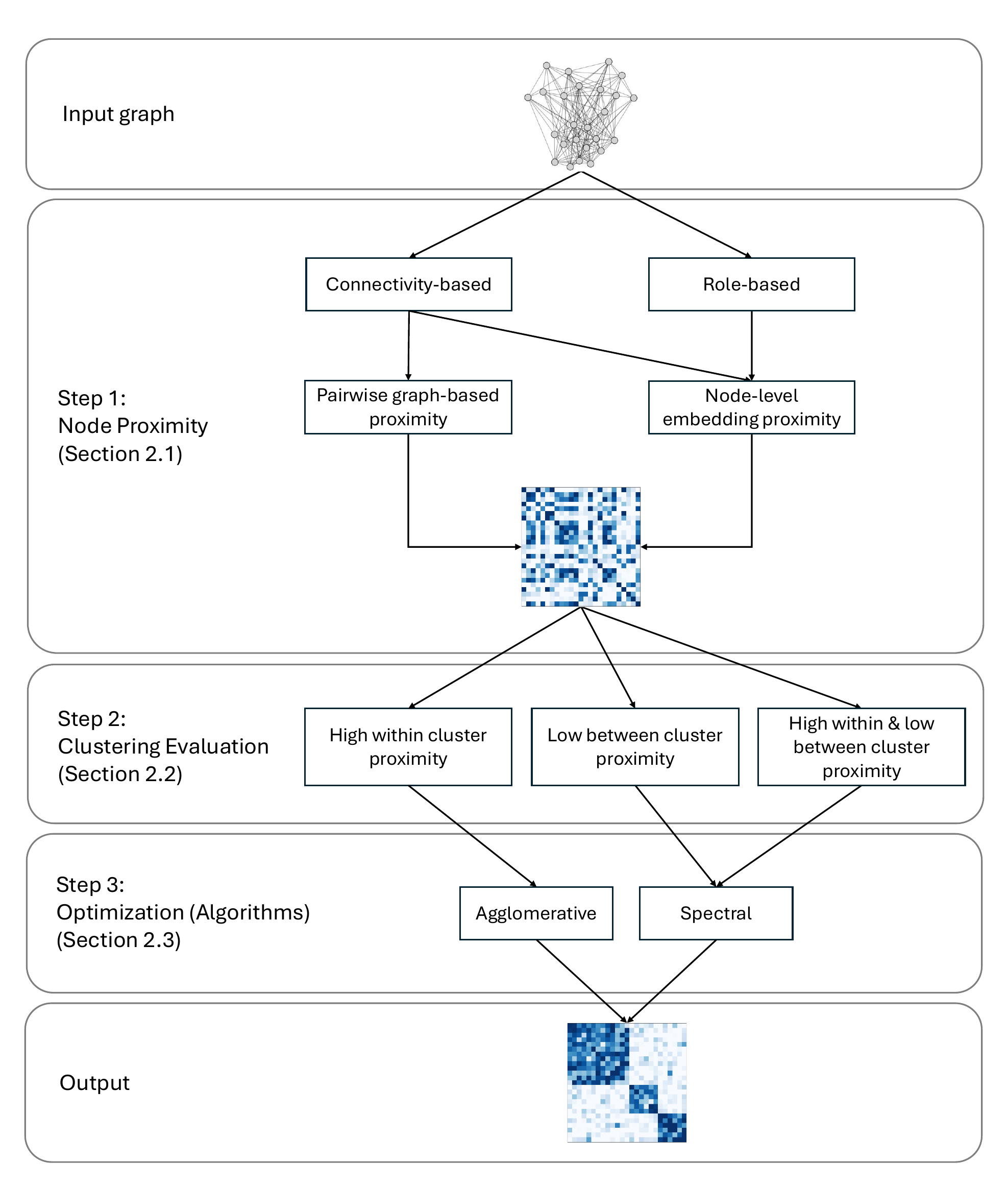}
    \caption{Flowchart of the three-stage clustering pipeline showing the various approaches to obtaining a clustering.}
    \label{fig:ovw}
\end{figure}

Instead of prescribing a singular clustering definition, this framework offers a flexible structure adaptable to various contexts and interpretations. Figure~\ref{fig:ovw} visually summarizes the conceptual structure of our clustering framework, highlighting the distinction between connectivity- and role-based approaches. It shows how proximity definitions, clustering objectives, and algorithmic choices align within a coherent pipeline, guiding the selection of appropriate methods depending on the modeler's context. More detailed explanations are provided in subsequent subsections. The section concludes by discussing alternative clustering methods and situating these methods within the perspective offered by the current framework.

In the following, let $G = (V,E)$ be a directed graph, with node set $V = \{1, \ldots, N\}$ and edge set $E \subset V \times V$ denoting the connections, where for now we simply focus on the single-layered case.
We can represent the graph using an adjacency matrix:
\begin{align*}
    A_{ij} = \begin{cases} 1, & \text{if there exists a directed edge } (i, j) \in E; \\ 0, & \text{otherwise}. \end{cases}
\end{align*}
Later in the paper, we extend this framework to the multi-layered case, where we explore how our approach generalizes to graphs with multiple layers.

\subsection{Step 1: Node Proximity } \label{sec:prox}

\begin{figure}[htbp]
    \centering
    \begin{subfigure}[b]{0.475\linewidth}
        \centering
        \includegraphics[width=\linewidth]{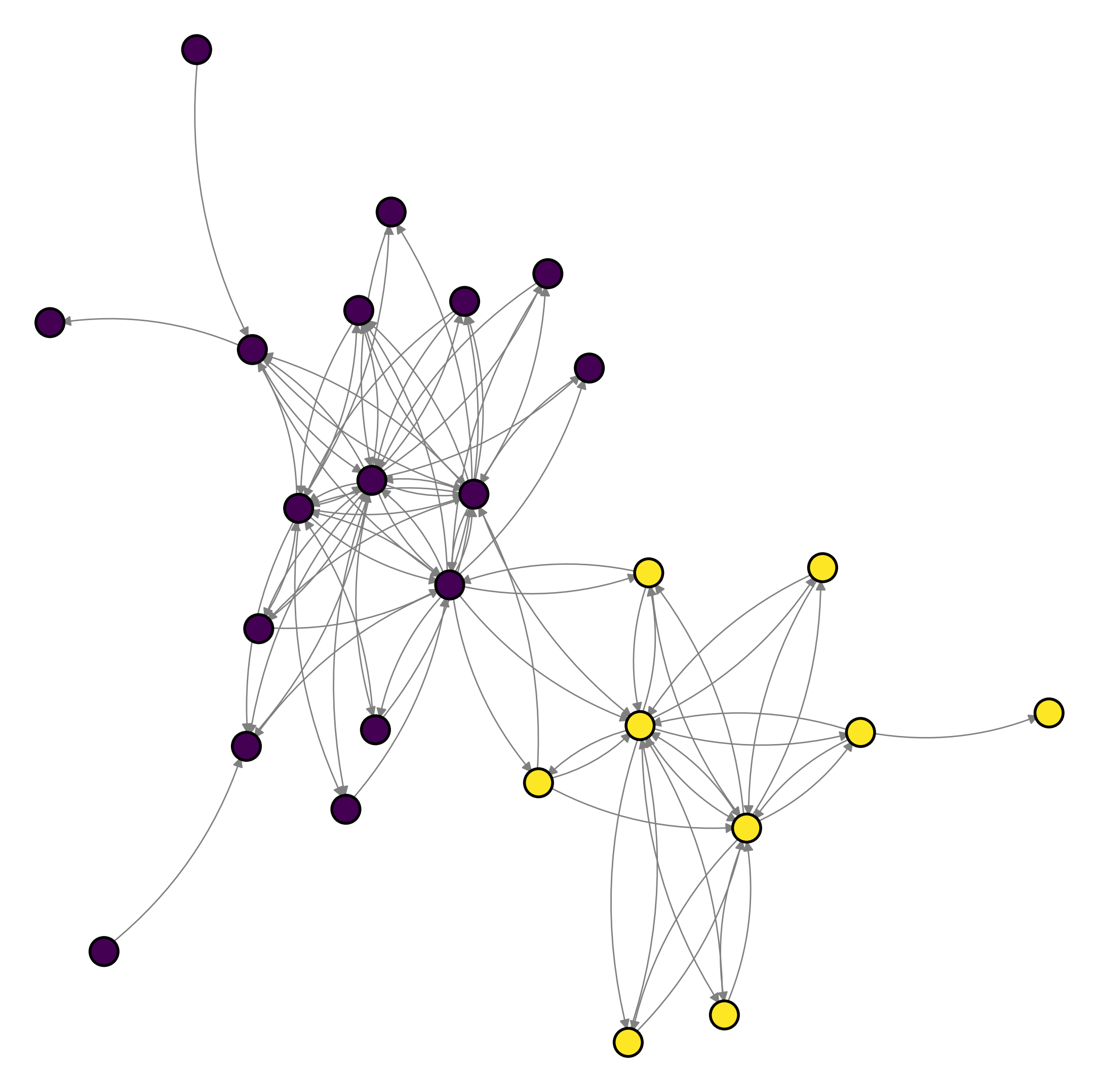}
        \label{fig:structural}
    \end{subfigure}
    \hfill
    \begin{subfigure}[b]{0.475\linewidth}
        \centering
        \includegraphics[width=\linewidth]{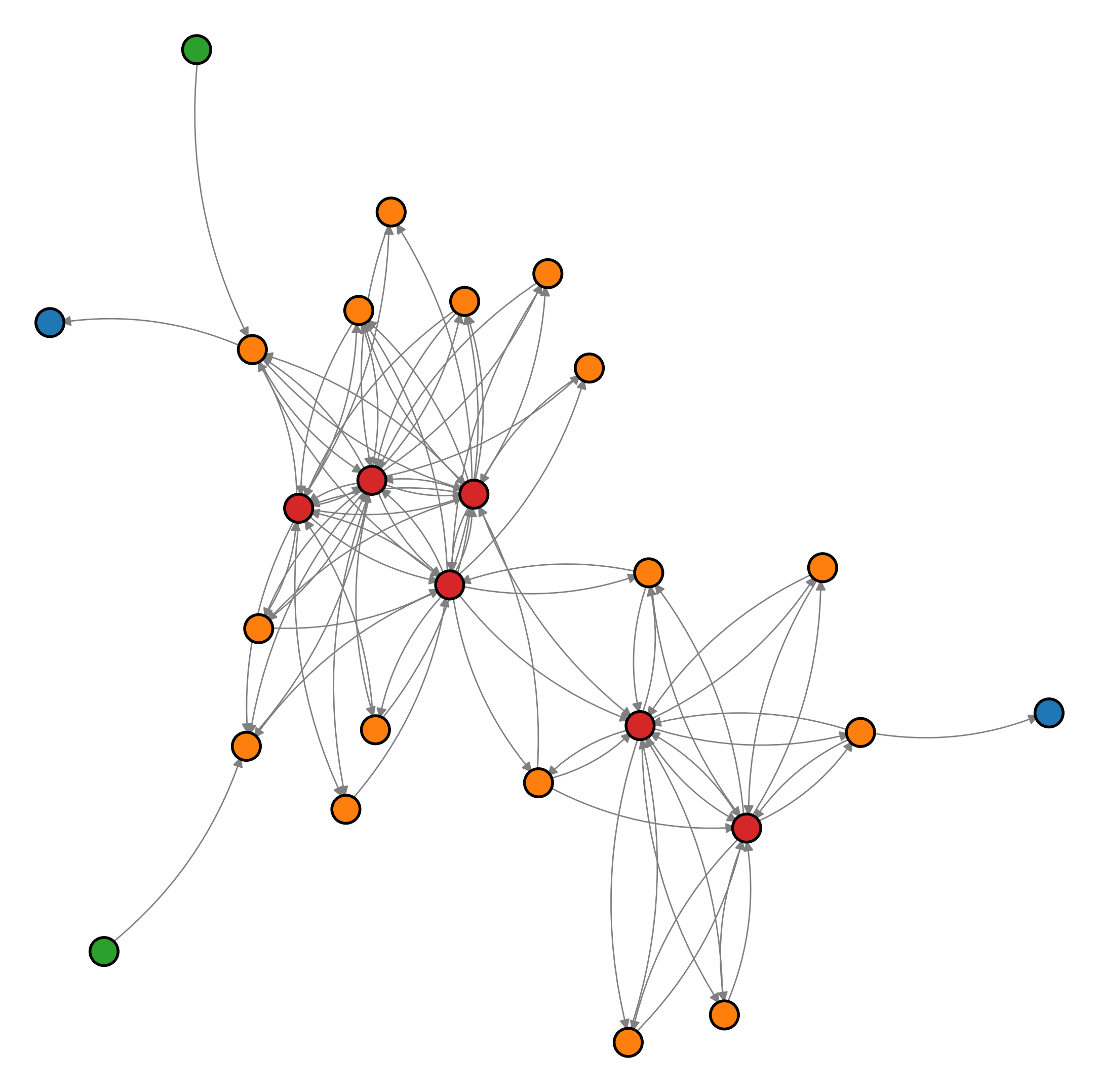}
        \label{fig:roll_based}
    \end{subfigure}
    \caption{The same graph clustered two ways: connectivity-based (left), grouping densely linked neighbors, versus role-based (right), grouping nodes by structural role (red core, orange first-tier periphery, green sending periphery, blue receiving periphery). Nodes sharing a role need not be directly connected.}
    \label{fig:rol_vs_com}
\end{figure}

Graph clustering of nodes can broadly be categorized into two domains: connectivity-based clustering and role-based clustering. Connectivity-based clustering focuses on grouping nodes that are \textit{proximate} in the graph's topology. For example, proximity can be measured via shortest path lengths.
In contrast, role-based clustering does not necessarily require nodes to be proximate in the graph itself; however, nodes must be \textit{similar} with respect to their functional role in the network. More specifically, the graph is used to construct a feature space, and clustering is performed in this space without being constrained by the graph's topological properties. We show an example in Figure~\ref{fig:rol_vs_com}, demonstrating a clear distinction between a connectivity-based clustered network (left panel) and the same network with a possible role-based clustering (right panel). Following the core-periphery model, one can identify a core (red), a first-tier periphery (orange), a second-tier sending periphery (green), and a second-tier receiving periphery (blue). Figure~\ref{fig:rol_vs_com} contrasts  outcomes of connectivity-based and role-based clustering on the same network, emphasizing that nodes occupying similar structural roles (right) may not be directly connected (as seen on the left).

Note that both concepts of proximity and similarity underpin the idea of clustering by grouping entities that are "close" to each other, either in a connectivity sense, where high connectivity between nodes means that the nodes are close in the graph's topology, or a role-based sense, where the two nodes are similar in behavior.
Clustering via proximity in terms of connectivity and role-based similarity can be unified within a single formal framework, where $S = (S_{ij})_{(i,j) \in V \times V}$, and $S_{ij} \in \mathbb{R}$ is the proximity or similarity between node $i$ and $j$ in graph $G$. In the remainder of the paper, we use proximity and similarity interchangeably, with distance treated as their antonym. Moreover, we set $S_{ii} = 0$ for all $i = 1, \ldots N$.

We next outline clustering methods based on pairwise graph-based or node-level embedding proximity. For a comprehensive overview of these and related approaches, we refer the reader to \citet{Fortunato_2010}. 

\subsubsection{Pairwise graph-based proximity}

Within the financial network setting, connectivity-based clustering can be of interest when the aim is to capture clusters of financial institutions from a financial contagion perspective. In this context, connectivity-based clustering helps uncover high-risk communities whose structural positions in the network make them particularly susceptible to, or capable of, transmitting financial distress. Recent work by \citet{torri2023financial} shows that community structures within banking networks can significantly amplify the spread of liquidity shocks, further emphasizing the importance of detecting such clustered topologies in systemic risk analysis.

When the aim is to group the graph's nodes based on connectivity, one can resort to a pairwise graph-based measure of distance between nodes.
Examples of literature utilizing such an approach include \citet{berkhout2019analysis}, who use the Kemeny constant to decompose a Markov influence graph. Moreover, \citet{yen2007graph} compute a proximity measure taking into account commuting times based on the graph's weighted adjacency matrix. Also, \citet{bartesaghi2020community} use the communicability distance introduced by \citet{estrada2008communicability} to minimize intra-cluster distance.

A practical consideration in using pairwise graph-based proximity is that the graphs we model are typically directed (e.g., financial transactions typically have a clear flow of assets). As a result, the distance from node $i$ to $j$ may not necessarily be the distance from node $j$ to $i$. However, clustering involves partitioning the node set into groups, and such partitions are inherently symmetric. Thus, while proximity may be asymmetric, clustering requires a symmetric comparison of nodes.

To address this asymmetry, it is common to work with a symmetrized proximity matrix. Thus, in case $S$ is not symmetrical, we can work with a symmetrized proximity matrix, where one can use: i) average pair-wise connectivity  $S^{sym}_{ij} = (S_{ij} + S_{ji}) / 2$; ii) minimal pair-wise connectivity $S^{sym}_{ij} = \min\{S_{ij}, S_{ji}\}$; or iii) maximal pair-wise connectivity $S^{sym}_{ij} = \max\{S_{ij}, S_{ji}\}$. 

\subsubsection{Node-level embedding proximity}

Another way of defining node proximity is through the construction of node embeddings. This approach can be used both in the case of connectivity-based clustering, as well as role-based clustering.

An example study utilizing a node-level embedding approach for connectivity-based clustering is \citet{zhou2003distance}, who construct a node embedding via the mean first passage time of a random walker on the graph. 
Moreover, \citet{Pecora_Kaltwasser_Spelta_2016} apply a connectivity-based clustering on e-MID data via a non-negative matrix factorization of the input graph.

Studies utilizing a node-level embedding for role-based clustering approach include
\citet{henderson2012rolx}, who construct node embeddings using so-called neighborhood and recursive features, explicitly modeling a role-based clustering approach. Additionally, \citet{ribeiro2017struc2vec} propose struc2vec, a framework specifically designed to capture structural identity by constructing a multilayer graph that encodes structural similarity, independent of node proximity. 
\citet{grover2016node2vec} aim to mix both connectivity-based and role-based clustering, through learning feature representations for nodes by simulating biased random walks to capture both local and global network structures.

In the following, we require two different steps for defining node proximity: feature extraction, and choosing a similarity metric.\\

\noindent \textit{Step 1a: Feature extraction } \\

\noindent Let \( \phi: A \rightarrow \mathbb{R}^{d \times N} \) be a feature mapping from the adjacency matrix $A$  to a \( d \)-dimensional embedding vector for nodes \( i \in V \), where typically \( d \ll N \). Moreover, let $\phi(i)$ refer to the embedding for node $i$.
Choosing a feature mapping $\phi$ is not a trivial preprocessing step but an explicit modeling decision: it determines which structural signals from the graph are preserved and which are suppressed. Rather than viewing 
the embedding as a black–box compressor of the adjacency matrix, we can treat it as the place where domain insight enters the model. By selecting which structural summaries populate each coordinate of our embedding, we decide what it means for two nodes to be “close.” In this sense, the embedding serves as a lens through which the graph is viewed. Therefore, choosing an appropriate feature embedding is of vital importance for a good clustering outcome. 

For example, from the perspective of role-based clustering, one should ask: ``what defines separate roles?''
If one wants to exclusively identify similar nodes in terms of supplying some good in a supply chain network, it can be sufficient to construct an embedding encoding the reachability from this node to other nodes (e.g. taking into account only out-degrees).
However, when grouping financial institutions to analyze the money market's functioning as a liquidity clearing market (we will discuss this market in more detail in Section~\ref{sec:fene}), one may want to identify similarity on the ability to both send and receive liquidity (e.g. taking into account both in- and out-degrees). 

A key advantage of such a white-box approach lies in its interpretable node embeddings, which provide immediate and meaningful role assignments. This sets it apart from existing node embedding methods, which typically do not incorporate handcrafted features when compressing a graph's structure. Instead, these methods, such as those proposed by \citet{grover2016node2vec} and \citet{hamilton2017inductive}, prioritize the learning of features that optimally compress the graph's structure without explicitly ensuring interpretability. 
As a result, the clusters identified may not necessarily correspond to meaningful roles in financial transaction networks. \\

\noindent \textit{Step 1b: Similarity metric } \\

\noindent To quantify the proximity between two node embeddings, we define a proximity function 
\(\sigma: \mathbb{R}^d \times \mathbb{R}^d \rightarrow \mathbb{R}\). Then, we construct the proximity (or similarity) matrix 
\(S \in \mathbb{R}^{N \times N}\), with entries given by pairwise proximities
\[
S_{ij} = \sigma\big(\phi(i), \phi(j)\big),
\]
for all \(i,j \in V \times V\). Different choices of the proximity function \(\sigma\) highlight different aspects of node proximity and can therefore influence the resulting clustering. A straightforward and interpretable distance-based approach is to compute the Minkowski distance with \(p = 1\), corresponding to the \(L_1\)-norm, resulting in the distances
\[
D_{ij} = \| \phi(i) - \phi(j) \|_1,
\]
for all \(i,j \in V\). Note that other choices for measuring distance can be used, such as the Mahalanobis distance. To obtain the corresponding similarity matrix $S$ from a distance matrix $D$, one can compute
\[
S_{ij} = \sigma(\phi(i), \phi(j)) = \max_{k,l} D_{k,l} - D_{ij},
\]
for all \(i,j \in V\). Other similarity measures, such as Pearson correlation or cosine similarity, may also be used depending on the properties of the embedding space and the desired notion of proximity.

\subsection{Step 2: Clustering Evaluation } \label{sec:ob}

The aim of network clustering approaches is to partition the set $V$ into $M$ different clusters $C_m \subset V$, $m=1,\ldots, M$.
We formalize this clustering problem via the grouping of similar nodes, the separation of dissimilar nodes, or both.
To that end, we make use of the similarity matrix $S$ and seek a permutation of the nodes such that the permuted matrix approximates a block-diagonal structure, where entries close to the diagonal exhibit high proximity and off-diagonal entries reflect low similarity. 
We illustrate such a transformation in Figure~\ref{fig:ovw}.

First, consider the \textit{within cluster similarity}
\begin{align}
    \Phi^{within}(S; C_1 \ldots, C_M) = \sum_{m =1}^M \frac{1}{\kappa(C_m)} \sum_{i,j \in C_m : j\neq i} S_{ij},
\end{align}
and its corresponding maximization to find a clustering:
\begin{align}  \label{eq:within}
    \max_{C_1, \ldots, C_M} \  \Phi^{within}(S; C_1 \ldots, C_M).
\end{align}
The normalization by $\kappa(C_m)$ is chosen to be either proportional to the cluster size, e.g., $\kappa(C_m) \propto |C_m|$ or to the cluster volume, e.g.,  $\kappa(C_m) \propto \sum_{i\in C_m} \sum_{j \in V : j\neq i} S_{ij}$. Note that the normalization is required to produce balanced clusters, i.e., clusters of somewhat comparable sizes. Without the normalization, maximizing \eqref{eq:within} is typically achieved by isolating a single node.

Second, one can quantify the \textit{between cluster similarities}:
\begin{align}
    \Phi^{between}(S; C_1 \ldots, C_M) =  \sum_{m =1}^M  \frac{1}{\kappa(C_m)} \sum_{i \in C_m} \sum_{j \in C_l : m \neq l} S_{ij}
\end{align}
and its corresponding minimization to find a clustering:
\begin{align}  \label{eq:between}
    \min_{C_1, \ldots, C_M} \  \Phi^{between}(S; C_1 \ldots, C_M).
\end{align}
Note that in the specific case where we choose $\kappa(C_m) = \sum_{i\in C_m} \sum_{j \in V : i\neq j} S_{ij}$, minimizing \eqref{eq:between} also maximizes \eqref{eq:within}, given that
\begin{align*}
     \sum_{i \in C_m} \sum_{j \in C_l :m \neq l} S_{ij} = 2 \sum_{i \in C_m} d_i(S) - \sum_{i,j \in C_m : j\neq i} S_{ij} = 2 \kappa(C_m) - \sum_{i,j \in C_m : j\neq i} S_{ij},
\end{align*}
where $d_i(S) = \sum_{j=1}^N S_{ij}$. Given that $M$ is constant, it follows that we \textit{jointly minimize between cluster similarity and maximize within cluster similarity}.

\subsubsection*{Choosing the number of clusters}

Choosing the most appropriate number of clusters is a generic problem in graph clustering. In this section, we argue that our objective can be used to optimize over the number of clusters $M$ via a grid search.
To that end, let $\Phi^{within}(S; C_1 \ldots, C_M) $ denote the quality of a graph's clustering.
Observe that for $\kappa(C_m) = \sum_{i\in C_m} \sum_{j \in V : j\neq i} S_{ij}$, we effectively compute the (normalized) within-similarity for each node in the graph.
Therefore, we aim to separate a single cluster describing our data into $M$ separate ones if  
\begin{align}
    \frac{1}{\kappa(C_m)}\sum_{i,j \in C_1 : j\neq i} S_{ij} < \sum_{m =1}^M \frac{1}{\kappa(C_m)}\sum_{i,j \in C_m : j\neq i} S_{ij},
\end{align}
that is to say that the (normalized) within-similarity per node increases if the cluster is separated into smaller clusters.
Now, we can use $\Phi^{within}(S; C_1 \ldots, C_M) $ to find an optimal number of clusters given $S$, that is we aim to find
\begin{align} \label{eq:opt_M}
     \max_{M \in \mathbb{N}} \max_{C_1, \ldots, C_M}  \Phi^{within}(S; C_1 \ldots, C_M).
\end{align}

\subsection{Step 3: Optimization Algorithms } \label{sec:alg}

Problems \eqref{eq:within} and \eqref{eq:between} are known to be NP-hard \citep{shi2000normalized}.  This complexity arises due to the combinatorial nature of assigning points to clusters. As a result, even moderately sized instances of these problems cannot be solved to optimality within a reasonable computational time. Therefore, we resort to approximations which we discuss in the following.

\subsubsection{Agglomerative clustering} 

Agglomerative clustering is a bottom-up hierarchical clustering approach that iteratively merges clusters based on their pairwise similarity or distance until all nodes form a single cluster. While similarity measures indicate closeness (higher similarity means closer clusters), agglomerative clustering typically operates using distances, where a smaller distance indicates higher similarity. Initially, each node represents its own cluster, and at each step, the two clusters with the smallest distance (equivalently, highest similarity) are merged according to a chosen linkage criterion, such as single, complete, or average linkage.

For instance, the average linkage method quantifies the average similarity between two clusters $m \neq m'$:
\begin{align*}
    \frac{1}{2|C_m||C_{m'}|} \sum_{i \in C_m, j\in C_{m'}} S_{ij},
\end{align*}
and merges the two clusters exhibiting the highest average similarity (or equivalently, minimal average distance).
Thus, in this specific setting, agglomerative clustering solves \eqref{eq:within} for $\kappa(C_m) = |C_m|^2 - |C_m|$ in a greedy approach.

\subsubsection{Spectral clustering} 

Spectral clustering is a technique that utilizes the eigenvalues of the similarity matrix $S$ to divide data into clusters, e.g., see \citet{von2007tutorial}. By computing the eigenvectors of the Laplacian $L = \mathcal{D} - S$, for diagonal matrix $\mathcal{D} = \text{diag}(S \Bar{1}) $ and vector of ones $\Bar{1}$, the data is embedded into a feature space, based on the largest eigenvectors of the Laplacian $L$. Then, conventional clustering methods like K-means are applied. Unlike agglomerative clustering, spectral clustering is not a greedy approach and considers the global structure of the similarity graph.
It is particularly useful for identifying complex, non-convex cluster structures. However, as opposed to hierarchical clustering, it does not necessarily find a hierarchy of clusters.

Spectral clustering solves a different objective depending on whether one normalizes the Laplacian or not. When using the normalized Laplacian, one effectively solves a relaxation of the Normalized Cut problem, that is \eqref{eq:between} for $\kappa(C_m) = \sum_{i\in C_m} \sum_{j \in V : j\neq i} S_{ij}$ \citep{von2007tutorial}. In contrast, using the unnormalized Laplacian leads spectral clustering to minimize a relaxation of the Ratio Cut objective, corresponding to $\kappa(C_m) = |C_m|$ in \eqref{eq:between}.
It can be shown that spectral clustering on the normalized Laplacian yields more balanced clusters in the number of nodes, than, for example, using spectral clustering on the regular Laplacian, or hierarchical clustering. The result is a less frequent occurrence of isolated nodes as clusters, which often is considered a desirable property when clustering.

\subsubsection{Alternative clustering methods} 

Many clustering methods do not directly align with the framework described. Unlike our approach, these methods typically do \textit{not} explicitly model proximity (see Figure~\ref{fig:ovw}). Instead, they define a clustering objective directly on the input graph and employ a suitable algorithm to optimize it \citep{newman2006modularity}.

For example, clustering algorithms like Louvain \citep{blondel2008fast} and Leiden \citep{traag2019louvain} iteratively optimize partitions to maximize modularity, a measure of the density of edges within clusters compared to a random graph. While these methods are efficient and effective at identifying densely connected regions, they are inherently tied to the graph's structural properties. 
The Leiden algorithm refines the Louvain method by introducing additional steps to ensure well-connected clusters, effectively including a constraint to \eqref{eq:within} to force clusters to be connected components, thereby improving the robustness of the results.
Moreover, flow-based methods, such as InfoMap \citep{rosvall2008maps}, treat clustering as an information-theoretic optimization problem. 
Flow-based methods aim to minimize the description length of random walks within clusters, effectively retaining flows and ensuring intra-cluster cohesion. Although they provide a distinct perspective by emphasizing flow retention, they remain constrained by the graph's topology.

\section{Role-based Clustering in Multi-Layer Financial Networks} \label{sec:rol_mul_fin}

In this section, we discuss how the aforementioned framework can be applied to identify the roles of financial institutions using role-based clustering.
Our multi-layer approach is designed to capture both micro-level interactions and macro-level structures within a financial network. 
A key advantage of this method is its ability to identify institutions that, while not dominant within any single market, serve as crucial bridges connecting multiple market segments. For example, certain institutions may link different market segments, thereby facilitating liquidity transmission across layers. This cross-layer perspective is especially important for identifying hidden channels of systemic risk and mapping potential contagion pathways that traditional, single-layer or binary core–periphery models may overlook, particularly during periods of market stress when inter-market dependencies become more pronounced.

Throughout this section, we consider the case of the MMSR data, which provides granular insights into unsecured and secured funding transactions among financial institutions. Introduced by the European Central Bank (ECB) in 2016, it collects detailed transaction-level data from major financial institutions across unsecured and secured lending, foreign exchange swaps, and overnight index swaps. Our aim is to cluster financial institutions that are active in the EU money market, and specifically identify financial institutions that intermediate between different market segments. By analyzing MMSR data, we can observe how financial institutions interact within short-term funding markets. For instance, some institutions may act as liquidity providers in the repo market while simultaneously serving as major borrowers in the unsecured segment. Such dual-market engagement positions them as critical intermediaries, facilitating liquidity transmission and influencing overall market stability.

\subsection{Input Graph}

Define $G = (V,E^{(1)}, \ldots, E^{(L)})$ as a multi-layer directed graph, with set of financial institutions $V$ and edge sets $E^{(l)}$, $1 \leq l \leq L$, denoting an indicator of a trading relationship between two parties in the $l$th layer.
Each layer $l$ can be represented using an adjacency matrix:
\begin{align*}
    A^{(l)}_{ij} = \begin{cases} 1, & \text{if there exists a directed edge } (i, j) \in E^{(l)}, \\ 0, & \text{otherwise}. \end{cases}
\end{align*}
The different layers can represent different financial markets.
For example, one layer can represent the unsecured money market, and another can represent the repo market, that is the secured money market.
Alternatively, we can separate a short term repo market and a long term repo market, to identify which financial institutions possibly intermediate these markets.
When identifying roles within this multilayer network, the aim is to identify patterns in financial institution's behavior within these markets.

Note that there exist various modeling choices when defining the adjacency matrices $A^{(l)}$, for each layer $l$, depending on the specific features of financial interactions one aims to capture. The most straightforward approach is to define a trading relationship between two parties based on the mere occurrence of at least one transaction within a specified time window. Nonetheless, one can choose to enrich this structure by incorporating further information, such as the frequency of transactions or the volume of trades.

\begin{figure}[htbp]
    \centering
    \label{fig:mmsr}
       \begin{subfigure}[b]{0.45\textwidth}
        \centering
        \includegraphics[width=\linewidth]{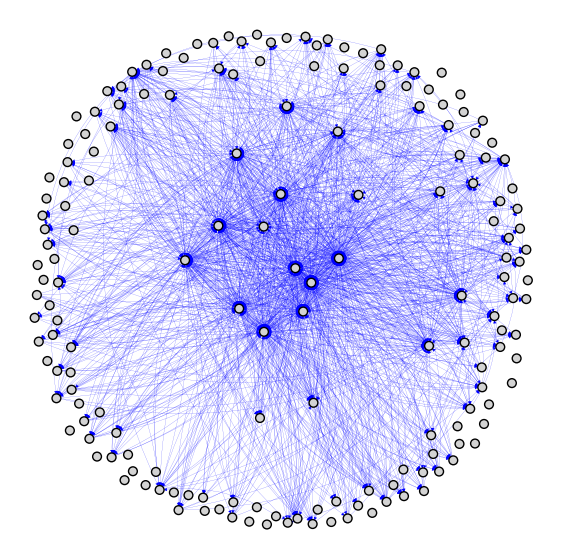}
        \label{fig:sec_net}
    \end{subfigure}
    \hfill
    \begin{subfigure}[b]{0.45\textwidth}
        \centering
        \includegraphics[width=\linewidth]{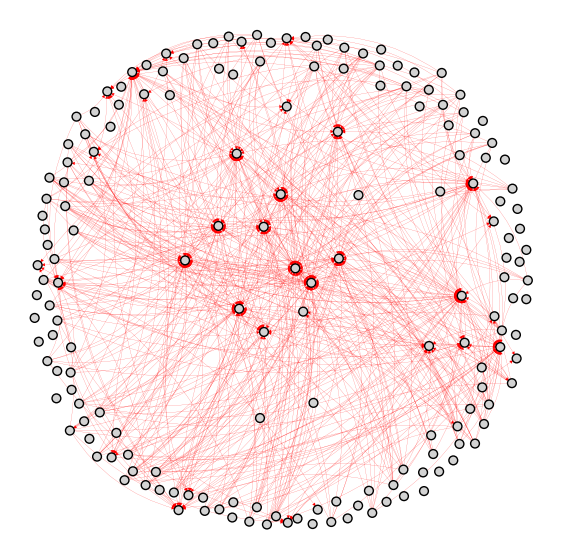}
        \label{fig:unsec_net}
    \end{subfigure}
    \caption{The 2023 MMSR network, with the secured layer (left, blue edges for repo lending) and unsecured layer (right, red edges for uncollateralized lending) shown separately. Both panels depict the same set of 173 banks in identical positions. This set is formed by a union of the top 100 largest banks in each market, measured by the number of counterparties.}
    \label{fig:sec_unsec_net}
\end{figure}

In Figure~\ref{fig:sec_unsec_net} we show a snapshot of the secured (blue) and unsecured (red) MMSR segments for the year 2023 (we plot the graph's layers separately for ease of plotting). Although we have access to the full data, we select the 100 largest banks, as measured by the number of counterparties, within each market segment to preserve confidentiality. We consider a financial institution active if that financial institution had at least one transaction during 2023 in either the (non-cleared) secured, or unsecured market segments. Moreover, we exclusively consider overnight transactions, and only select trades where the trade date is equal to the settlement date.
A red edge from $i$ to $j$ indicates that $i$ lent cash to $j$ at some point in 2023 without receiving collateral. A blue edge indicates that $i$ lent cash to $j$ in a repo transaction, thus receiving collateral in return. For the sake of this numerical example, we exclude the foreign exchange and overnight index swap segments from the MMSR data.

\subsection{Step 1: Node Proximity} \label{sec:fene}

In this application, we aim to group financial institutions based on their ability to trade with counterparties, potentially across multiple markets. A crucial feature of financial transactions, however, is that they often extend beyond direct trading relationships; indirect trading via intermediaries can also be important, particularly through mechanisms such as collateral reuse. As highlighted by \citet{inhoffen2024safe}, the reuse of collateral in the European repo market extends the effective supply of safe assets, allowing institutions to engage in funding transactions beyond their direct counterparties. This means that a financial institution's access to liquidity and market influence is not solely determined by its immediate trading partners but also by its position within longer transaction chains. Similarly, the pass-through of liquidity from the central bank through the system also involves chains of interactions.
By incorporating these indirect trading linkages into our role-based clustering framework, we can better identify institutions that serve as key nodes in the transmission of liquidity and risk across financial markets. 

In the following, we set forth an approach for capturing direct and indirect trading opportunities in a node embedding.
An effective method is through the analysis of egonets on the network of past transactions, which serve as localized subgraphs capturing both immediate and extended trading relationships \citep{wu2015egoslider}. 
By examining the patterns and relationships within egonets, it is possible to group nodes based on their functional similarities, highlighting functional categories. 

To embed the structure of each egonet into a feature representation, we adopt and extend the methodology introduced by \citet{beguerisse2013finding}, who originally proposed using egonet-derived features for single-layer networks based on path counts. We build upon this idea by introducing several modifications, most notably by extending it to handle multi-layer networks, where each layer corresponds to a different market. This extension enables the embedding to incorporate a richer, more comprehensive view of an institution's position within the financial network.
    
Let $A^{(l)}$ be the network of trading relationships in market $l$.
We then define the embedding through the number of direct and indirect (via an intermediary) connections a financial institution has (\textbf{Step 1a}).
To that end, we count the total $k$th order paths for each node, for $k=1,2,\ldots,K$;
\begin{align} 
\begin{split}
     \phi_{in}^{(l)} &= \left(\Bar{1}^{\top} A^{(l) } , \Bar{1}^{\top} \left( A^{(l)} \right)^2, \Bar{1}^{\top} \left( A^{(l)} \right)^3, \dots \right)^{\top}, \\
    \phi_{out}^{(l)} &= \left(  A^{(l)} \Bar{1}, \left( A^{(l)} \right)^2 \Bar{1}, \left( A^{(l)} \right)^3 \Bar{1}, \dots \right)^{\top}.
\end{split}
\end{align}
Note that there is a large serial correlation between features in each layer, since a path of length $k$ implies the existence of a path of length $k-1$.
To remove this correlation, 
we ensure that for path lengths $k \geq 2$, we divide by the number of paths of length $k-1$, that is 
\begin{align} \label{eq:phi_bar}
\begin{split}
     \Bar{\phi}_{in}^{(l)} &= \left(\Bar{1}^{\top} A^{(l) } , \Bar{1}^{\top} \left( A^{(l)} \right)^2 \oslash \Bar{1}^{\top} A^{(l) }, \Bar{1}^{\top} \left( A^{(l)} \right)^3 \oslash \Bar{1}^{\top} \left( A^{(l)} \right)^2, \dots \right)^{\top},  \\
    \Bar{\phi}_{out}^{(l)} &= \left(  A^{(l)} \Bar{1}, \left( A^{(l)} \right)^2 \Bar{1} \oslash A^{(l)} \Bar{1}, \left( A^{(l)} \right)^3 \Bar{1} \oslash  \left( A^{(l)} \right)^2, \dots \right)^{\top},
\end{split}
\end{align}
where $\oslash$ is the element-wise division.
The division ensures that we effectively count the average number of paths of length $k$ per path of length $k-1$, thus omitting any serial correlation.
Then, $\Bar{\phi}^{(l)} = (\Bar{\phi}_{in}^{(l)} \ || \ \Bar{\phi}_{out}^{(l)})$ denotes the feature matrix for each layer $l = 1\, \dots, L$, where $(\Bar{\phi}_{in}^{(l)} \ || \ \Bar{\phi}_{out}^{(l)})$ is the concatenation between a feature matrix for ingoing paths $\Bar{\phi}_{in}^{(l)}$, and outgoing paths $\Bar{\phi}_{out}^{(l)} $.
It follows that $\Bar{\phi}^{(l)}$ is an embedding for the egonets of all nodes for depth $K$ in the $l$th layer and in a clustering, the aim is to group the nodes based on their possibilities to trade.

\begin{figure}[bp]
    \centering
    \includegraphics[width=0.999\linewidth]{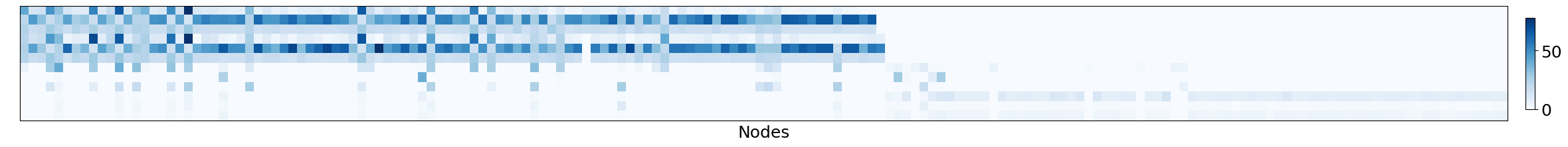}
    \caption{Heatmap of the egonet embedding matrix $\Bar{\phi} = (\Bar{\phi}_{in}^{(1)} \ || \ \Bar{\phi}_{out}^{(1)} \ || \Bar{\phi}_{in}^{(2)} \ || \Bar{\phi}_{out}^{(2)})$. Each column contains the embedding of a single bank and rows show in/out path-count features at depths 1, 2, and 3 for each layer.}
    \label{fig:embed}
\end{figure}

To compute an embedding for each node in Figure~\ref{fig:sec_unsec_net}, we compute features $\Bar{\phi}^{(l)}(i)$, for all nodes $i\in V$, for both the secured market segment (blue, $l=1$), and unsecured (red, $l=2$) market segment, up until path lengths $K=3$ (longer transaction chains become increasingly unlikely).
We show the embeddings in Figure~\ref{fig:embed}, where the features are represented as a matrix of column embeddings.

So far, we only count paths that start in one layer and remain within that same layer throughout. However, it is also possible to count paths that traverse across different layers, so as to capture the extent to which institutions can transact with counterparties that are intermediaries between different markets.

While it seems natural to count the number of paths for each node as a proxy for the trading possibilities of a financial institution, an alternative approach would be to count the number of distinct counterparties that are reachable within a given number of steps. This variation offers a more partner-oriented view of market accessibility, which may be particularly relevant when the diversity of trading relationships is of greater interest than the volume of indirect connections. We propose the path-counting method due to its simplicity, but we emphasize that the framework is flexible and can accommodate different notions of proximity or network reachability, depending on the specific analytical goals.

\begin{figure}
    \centering
    \begin{subfigure}[b]{0.49\linewidth}
        \centering
        \includegraphics[width=\linewidth]{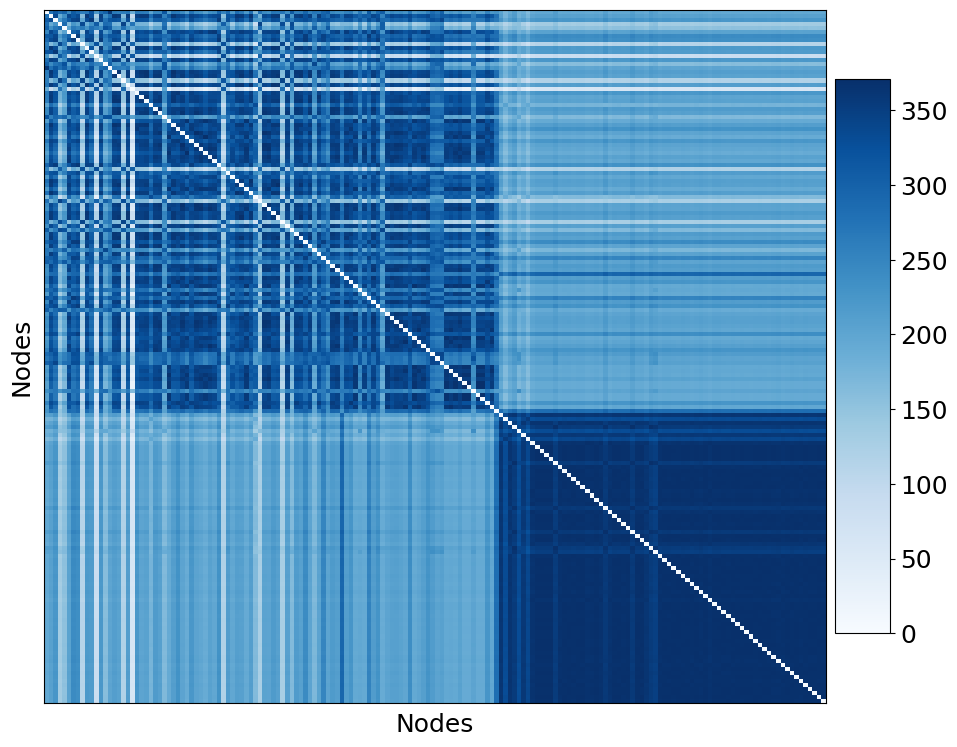}
        \caption{Similarity matrix of features shown in Figure~\ref{fig:embed}. \\
        \hspace{1pt}}
    \label{fig:sim_unclust}
    \end{subfigure}
    \hfill
    \begin{subfigure}[b]{0.49\linewidth}
        \centering
        \includegraphics[width=\linewidth]{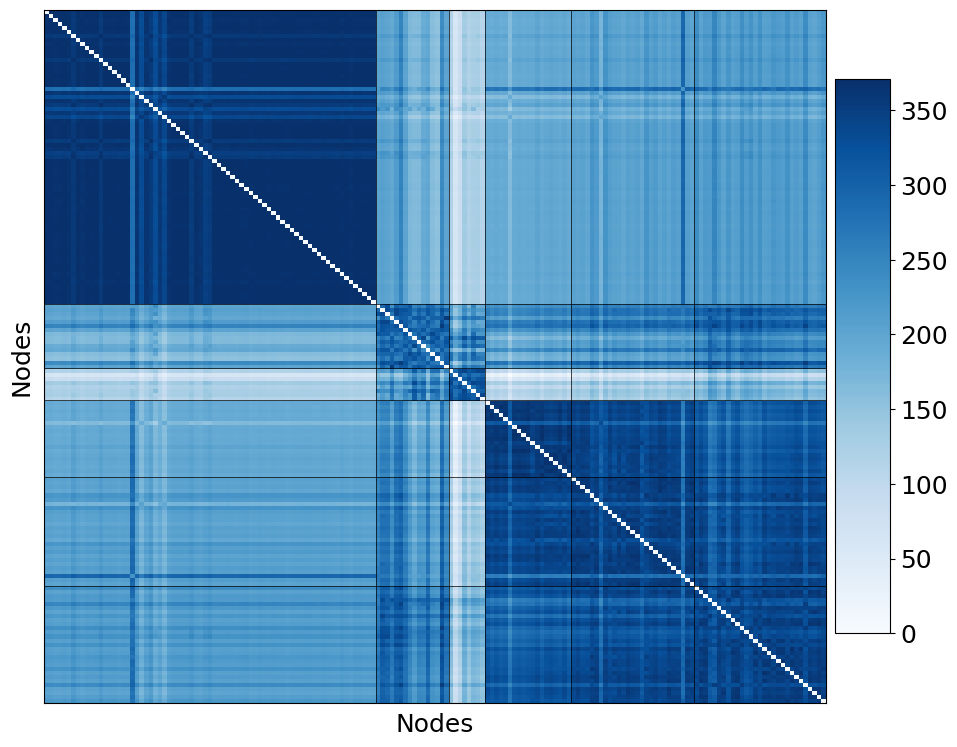}
        \caption{Clustered similarity matrix from Figure~\ref{fig:sim_unclust} using spectral clustering with $M=6$ clusters.}
    \label{fig:sim_clust}
    \end{subfigure}
    \caption{Pairwise similarity matrix $S$ shown (a) in the original node ordering and (b) permuted to emphasize the block-diagonal structure that results from the clustering pipeline.}
\end{figure}

Continuing with \textbf{Step 1b }for defining node proximity, we construct the similarity matrix. We let $S = D_{max} - D$, where $D_{max}$ is the maximum element of the Minkowski distance matrix $D$, where $p=1$. Thus, nodes with zero Minkowski distance have similarity $D_{max}$, and nodes with maximal Minkowski distance have similarity 0. We show the resulting similarity matrix in Figure~\ref{fig:sim_unclust} for the MMSR data in Figure~\ref{fig:sec_unsec_net}.

\subsection{Steps 2 \& 3: Clustering Evaluation \& Algorithm}

\begin{figure}
    \centering
    \includegraphics[width=0.65\linewidth]{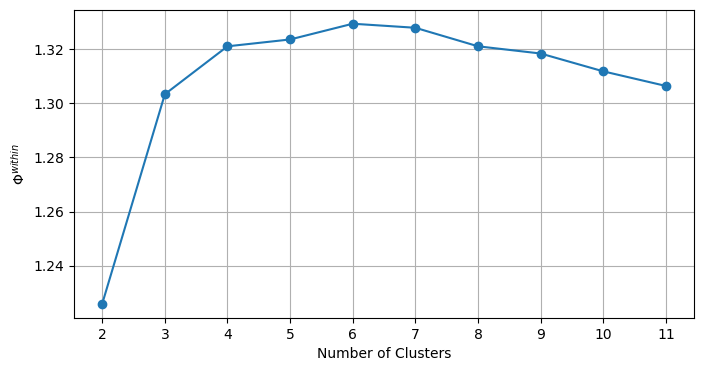}
    \captionof{figure}{Within-cluster similarity $\Phi^{within}$ versus the number of clusters $M = 2, \ldots, 11$ on MMSR data. The objective peaks at $M = 6$, indicating the optimal number of clusters is 6.}
    \label{fig:elbow}
\end{figure}

Having defined a similarity matrix from the embeddings, we now find a node clustering via maximizing within-cluster similarity and minimizing between-cluster similarity (\textbf{Step 2}). To that end, we solve \eqref{eq:opt_M} for $\kappa(C_m) = \sum_{i\in C_m} \sum_{j \in V : j\neq i} S_{ij}$ via spectral clustering so that we effectively solve \eqref{eq:within} (\textbf{Step 3}).
As discussed in Section~\ref{sec:alg}, this problem is NP-hard. Therefore, for each $M=2,\ldots, 11$, we perform spectral clustering with 500 different initializations for the centroids in K-means and choose the clustering that maximizes the objective value $\Phi^{within}(S; C_1 \ldots, C_M)$. 
We plot the results in Figure~\ref{fig:elbow}, where $M = 6$ shows an optimal clustering evaluation. Moreover, we show the associated clustered similarity matrix in Figure~\ref{fig:sim_clust}. 

\subsection{Clustering Results \& Interpretation} \label{sec:res}

\begin{figure}
    \centering
    \begin{turn}{90}
    \includegraphics[width=0.7\linewidth]{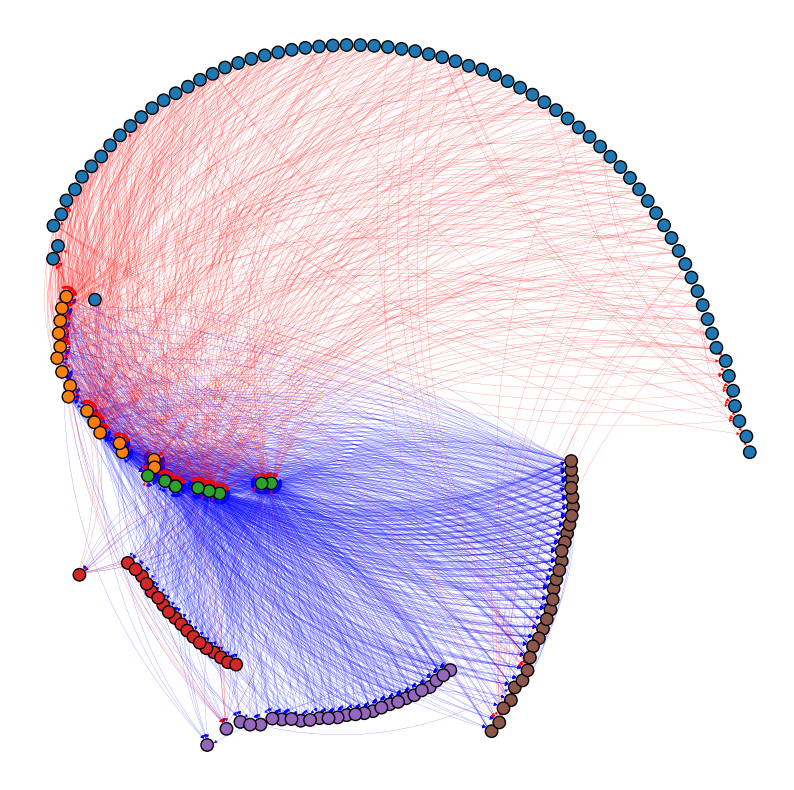}
    \end{turn}
    \captionof{figure}{Optimal role-based clustering for the network in  Figure~\ref{fig:sec_unsec_net},  with the 173 banks (nodes) colored by cluster (red, green, purple, brown, orange, blue). Blue edges denote secured transactions, red edges denote unsecured transactions, and outgoing edges curve counter-clockwise.}
    \label{fig:mmsr_clust}
\end{figure}

Having obtained the optimal clustering for $M= 6$ clusters in the previous step, we now illustrate the associated role-based clustering through a coloring of the nodes in Figure~\ref{fig:mmsr_clust}, where we plot the two segments in Figure~\ref{fig:sec_unsec_net} in a single graph. 

\begin{table}
    \centering
    \caption{The dimensions used to read each cluster's polar profile in Figure \ref{fig:mmsr_prof}: connectivity, funding balance, segment balance, and counterparty access (direct versus indirect links, distinguishing direct, semi-direct, and indirect reach). All readings are relative to the other clusters rather than absolute.} 
    \label{tab:reading_dimensions}
    \begin{tabular}{p{3cm} p{4.25cm} p{7.75cm}}
        \toprule
        \textbf{Dimension} & \textbf{Measurement} & \textbf{Economic meaning} \\ \midrule
        \textbf{Connectivity} & Sum of all elements in the embedding. & High values signal \textbf{large} or  \textbf{systemic} institutions, mid-range values \textbf{mid-tier} players, and low values \textbf{niche} or \textbf{peripheral} participants.\\[30pt]
        \textbf{Funding balance} & Total outgoing vs.\ ingoing connections. & Many outgoing connections indicate a cash \textbf{supplier}, many ingoing connections indicate a cash \textbf{consumer}. Dispersion indicates \textbf{intermediation}.\\[30pt]
        \textbf{Segment balance} & Total secured vs.\ unsecured connections.  & Concentration indicates \textbf{segment specialism}, dispersion indicates \textbf{cross-segment activity}.\\[17pt]
        \textbf{Counterparty access} & Total of direct connections vs. indirect connections. & A high share of direct links implies \textbf{direct counterparty access}. Conversely, a predominance of indirect links implies \textbf{indirect counterparty access}. A more balanced mix signals \textbf{semi-direct counterparty access}. \\ \bottomrule
    \end{tabular}
\end{table}

\begin{figure}
    \centering
    \begin{subfigure}[b]{0.475\textwidth}
        \centering
        \includegraphics[width=\linewidth]{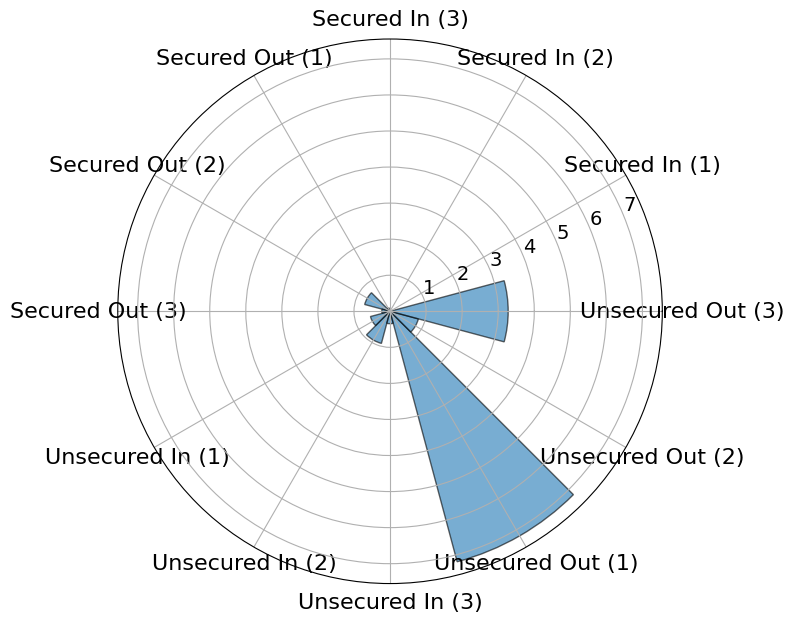}
        \label{fig:subfig1}
    \end{subfigure}
    \hfill
    \begin{subfigure}[b]{0.475\textwidth}
        \centering
        \includegraphics[width=\linewidth]{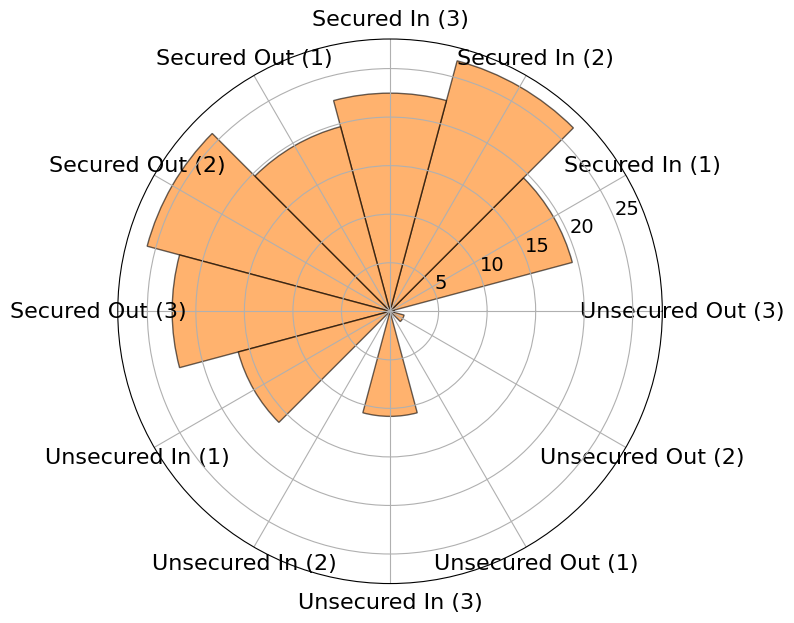}
        \label{fig:subfig2}
    \end{subfigure}

    \vspace{0.05cm}

    \begin{subfigure}[b]{0.475\textwidth}
        \centering
        \includegraphics[width=\linewidth]{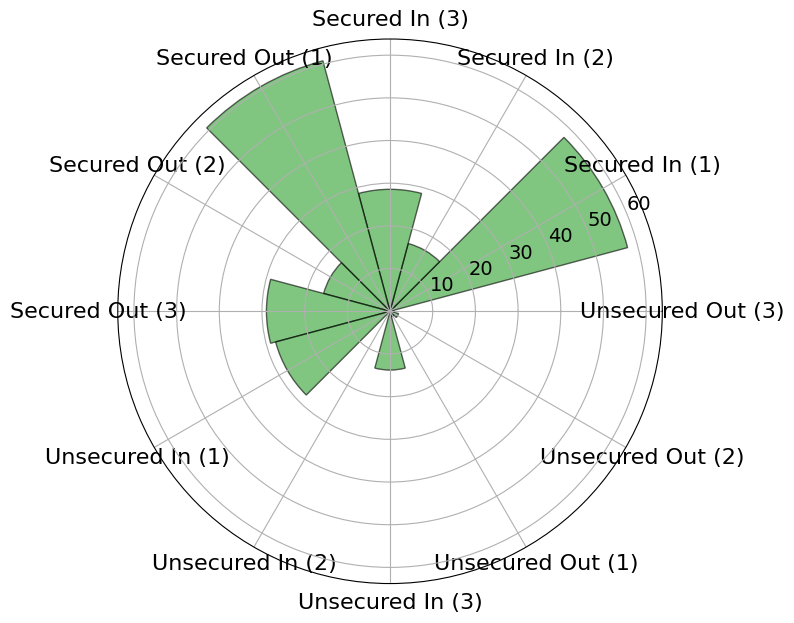}
        \label{fig:subfig3}
    \end{subfigure}
    \hfill
    \begin{subfigure}[b]{0.475\textwidth}
        \centering
        \includegraphics[width=\linewidth]{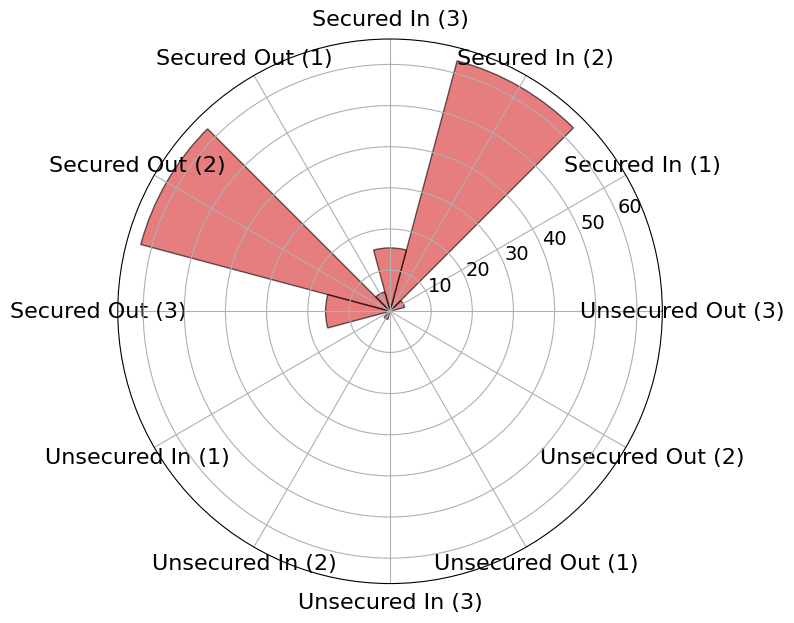}
        \label{fig:subfig4}
    \end{subfigure}

    \vspace{0.05cm}

    \begin{subfigure}[b]{0.475\textwidth}
        \centering
        \includegraphics[width=\linewidth]{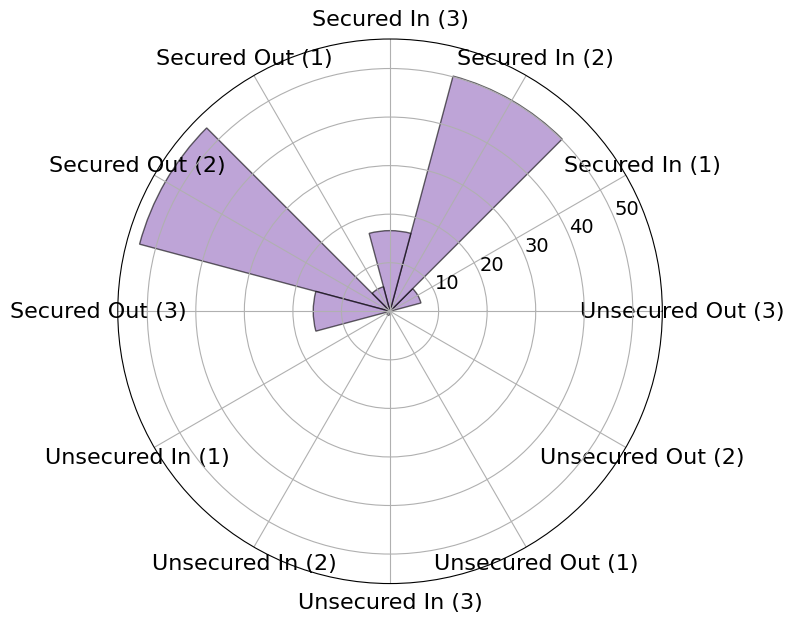}
        \label{fig:subfig5}
    \end{subfigure}
    \hfill
    \begin{subfigure}[b]{0.475\textwidth}
        \centering
        \includegraphics[width=\linewidth]{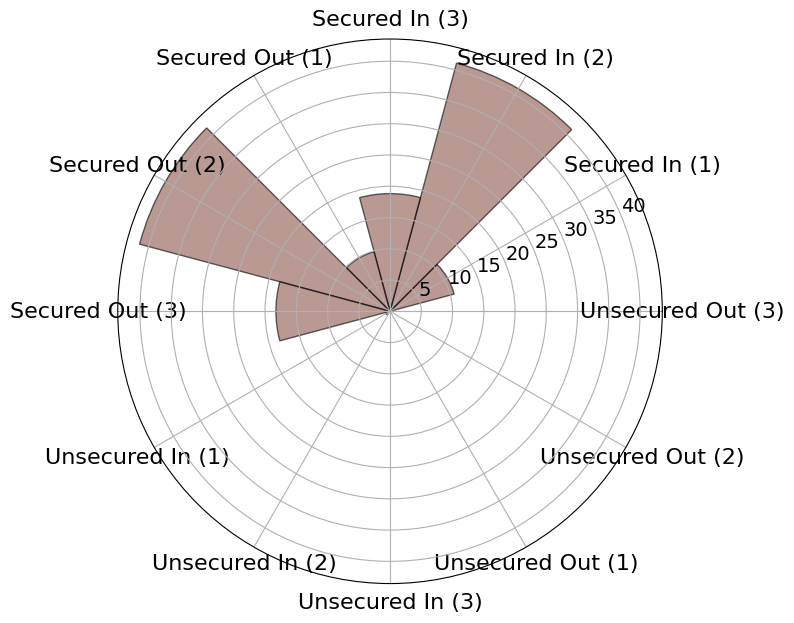}
        \label{fig:subfig6}
    \end{subfigure}
    \caption{Cluster-mean egonet profiles, one polar bar chart per cluster (colors match Figure~\ref{fig:mmsr_clust}). Each chart has twelve spokes: secured and unsecured markets, ingoing (In) and outgoing (Out) directions, at path lengths 1, 2, and 3.}
    \label{fig:mmsr_prof}
\end{figure}

To facilitate interpretation, we visualize the average embedding of nodes within each cluster in Figure~\ref{fig:mmsr_prof}. This plot reveals clear structural differences across clusters with respect to their relative market activity. As demonstrated below, these embeddings can be analyzed along four key dimensions, outlined in Table~\ref{tab:reading_dimensions}. These dimensions provide a basis for understanding the relative role of each cluster. A summary of these characterizations is presented in Table~\ref{tab:cluster_roles}. 
Appendix~\ref{app:missing_robust} confirms that these role assignments remain robust when we randomly remove edges to simulate incomplete reporting.

Inspecting the embedding of the green cluster, we observe many connections compared with the other profiles, indicating the green cluster contains systemically relevant institutions. Also, we see relatively equal outgoing compared to ingoing connections in the secured market, showing secured market intermediation.
However, in the unsecured market, we predominantly see cash borrowing. Moreover, since we observe activity
in both the secured and unsecured markets, the green cluster is cross-market intermediating. Finally, there is a relatively high number of direct connections, allowing the green cluster immediate counterparty access.

The blue cluster shows by far the lowest overall connectivity, marking it as a peripheral participant. It is active only in the unsecured segment and almost exclusively through outgoing links, behaving primarily as a cash supplier. Its absence from the secured segment gives an unsecured-market specialism, and because its links are predominantly first-order, its few counterparties are accessed directly.

The orange cluster also acts as a bridge between the secured and unsecured segments. Its connectivity is below
that of the red cluster but substantially above the peripheral blue cluster, thus indicating mid-tier connectivity.
The presence of material link weight in both segments yields a balanced segment-balance profile, and the mix
of outgoing and ingoing links in the secured market suggests market intermediation. Because a substantial share of these links are direct, its counterparty access is more direct than in the purple or blue clusters, yet higher than in the red cluster.

The purple cluster's connectivity is similar to the orange cluster's connectivity. It is only active in the secured money market marking the cluster as a segment specialist. The secured-segment links are broadly balanced between outgoing
and ingoing, consistent with intermediation. 
The large share of second-degree over first-degree links persists, implying reliance on indirect counterparties.

The red cluster is similar to the purple cluster with a subtle difference that there is less direct counterparty access and a stronger indirect counterparty reach.

Finally, the brown cluster has a slightly lower total link weight than the purple and red clusters but remains mid-tier when compared with the peripheral blue cluster. It is also exclusively active in the secured segment, where it intermediates with a roughly even split between lending and borrowing.

\begin{table}[h]
    \centering
    \caption{Qualitative role assignments for the six clusters along the four dimensions of Table~\ref{tab:reading_dimensions}, summarising the interpretation of Figure \ref{fig:mmsr_prof}}
    \label{tab:cluster_roles}
    \begin{tabular}{lllll}
        \toprule
        \textbf{Cluster} & \textbf{Connectivity} & \textbf{Funding balance} & \textbf{Segment balance} & \textbf{Counterparty access} \\ \midrule
        Green    & Systemic    & Intermediation & Cross-segment activity   & Direct \\[4pt]
        Orange  & Systemic    & Intermediation & Cross-segment activity      & Semi-direct \\[4pt]
        Purple & Mid-tier    & Intermediation & Segment specialism (sec.)     & Semi-direct  \\[4pt]
        Red & Mid-tier    & Intermediation & Segment specialism (sec.)   & Indirect \\[4pt]
        Brown  & Mid-tier    & Intermediation & Segment specialism (sec.)   & Semi-direct \\[4pt]
        Blue   & Peripheral  & Supplier       & Segment specialism (unsec.) & Direct\\ \bottomrule
    \end{tabular}
\end{table}

\section{Discussion} \label{sec:conc}

This paper provides both a practical and interpretable framework for role-based clustering in multi-layer financial networks. We outline a general clustering framework, centered on proximity definitions of nodes, clustering objectives, and optimization algorithms, that is adaptable to a wide range of contexts. Building on this foundation, we propose a role-based approach using egonet-based embeddings to uncover functional positions of institutions across different market layers. The empirical illustration using MMSR data highlights how this method can identify institutions that may be crucial for liquidity transmission and systemic stability, even if they are not highly connected within individual segments. 

While traditional analyses of financial networks often begin with predefined structures, such as the core–periphery model \citet{CRAIG2014322}, our approach avoids such assumptions. Rather than imposing a classification of institutions as either core or peripheral, we adopt a model-free framework that allows institutional roles to emerge from the data itself. By leveraging explainable node embeddings based on local network features (egonets), we uncover a variety of institutional roles across money market segments, such as systemic and mid-tier intermediaries, secured-market specialists, unsecured-market suppliers, and cross-segment bridges. Such roles are not only shaped by their overall connectivity, but also by their position within and across market layers, offering a more flexible and interpretable perspective on intermediation in the financial system.

Future research offers ample opportunities to further develop and extend this approach. Potential extensions include incorporating transaction frequencies or trade values into the embedding process, exploring alternative embedding strategies, such as counting distinct counterparties reachable within given steps, and expanding the analysis to include additional market segments like short-term versus longer-term funding.
Beyond methodological improvements, the framework holds considerable promise for practical applications. Firstly, a full-scale empirical application to MMSR data, distinguishing additional market segments, could provide deeper insights into the robustness and broader applicability of the proposed framework.
Moreover, tracking institutional roles dynamically over time, as also done by \citet{Kojaku_2018}, can support real-time monitoring of market structure, helping supervisors detect early signs of systemic instability and shifts in intermediation patterns.
Finally, we could examine whether particular roles come with tangible cost or benefits, for example, in terms of funding costs or liquidity access. 

\section*{Acknowledgements}

The authors would like to thank Kartik Anand, Cars Hommes, Tiziano Squartini, for their time to provide thoughtful and constructive feedback on earlier versions of this manuscript. Their insights and suggestions greatly contributed to improving the clarity and quality of the work. The views expressed in this paper are solely those of the authors and do not represent the official views of De Nederlandsche Bank.

\clearpage
\bibliography{References.bib}

\clearpage
\appendix
\addcontentsline{toc}{section}{Appendix}

\section{Robustness to missing links} \label{app:missing_robust}
Transaction-level supervisory data may contain some degree of incomplete reporting \citep{anand2018missing}. This appendix examines how sensitive the role-based clustering of Section~\ref{sec:rol_mul_fin} is to such missingness in the following scenario. Edges are removed independently and uniformly at random within each segment, representing unsystematic reporting gaps. We remove a fraction $p \in \{0.05, 0.10, 0.20, 0.30, 0.50\}$ of edges and re-run the full clustering pipeline, holding $M=6$ fixed throughout to facilitate a comparison with the baseline clusters achieved in Section~\ref{sec:res} (i.e., $p=0$). For each $p$ we perform 20 independent replications with distinct random seeds. 

We report two complementary diagnostics. The first is the adjusted Rand index (ARI), which measures the agreement of the whole partition with the baseline. It counts node pairs assigned consistently across the two clusterings, corrected for chance, and equals $1$ for identical partitions and approximately $0$ for partitions no more concordant than random labellings. Let $\mathcal{U}$ and $\mathcal{V}$ be two partitions of the $N$ nodes, let $n_{ij} = |U_i \cap V_j|$ count the nodes assigned to cluster $U_i$ under one partition and $V_j$ under the other, and let $a_i = \sum_j n_{ij}$ and $b_j = \sum_i n_{ij}$ denote the cluster sizes. The ARI is
\begin{align} \label{eq:ari}
    \mathrm{ARI} = \frac{\sum_{i,j}\binom{n_{ij}}{2} - \mathbb{E}}{\tfrac{1}{2}\left[\sum_i\binom{a_i}{2} + \sum_j\binom{b_j}{2}\right] - \mathbb{E}}, \qquad \mathbb{E} = \frac{\sum_i\binom{a_i}{2}\,\sum_j\binom{b_j}{2}}{\binom{N}{2}},
\end{align}
where $\mathbb{E}$ is the value of $\sum_{i,j}\binom{n_{ij}}{2}$ expected under random partitions with the same cluster sizes. 

The second diagnostic is the per-cluster mean $F_1$ score, which measures how reliably each individual role is recovered: treating the baseline assignment of a cluster as ground truth and the perturbed assignment as a prediction,
\begin{align} \label{eq:f1}
    F_1 = \frac{2\,\mathrm{TP}}{2\,\mathrm{TP} + \mathrm{FP} + \mathrm{FN}},
\end{align}
where a node is a true positive (TP) if assigned to the cluster under both partitions, a false positive (FP) if only under the perturbed one, and a false negative (FN) if only under the baseline. It equals one when the two assignments coincide and falls as members are gained or lost. Because cluster labels are arbitrary, $F_1$ first requires matching each perturbed cluster to its baseline counterpart. We obtain this matching by maximising total overlap over all one-to-one label correspondences, an assignment problem solved exactly by the Hungarian algorithm \citep{kuhn1955hungarian}, and evaluate \eqref{eq:f1} along the matched pairs.

\begin{figure}[htbp]
    \centering
    \includegraphics[width=0.6\linewidth]{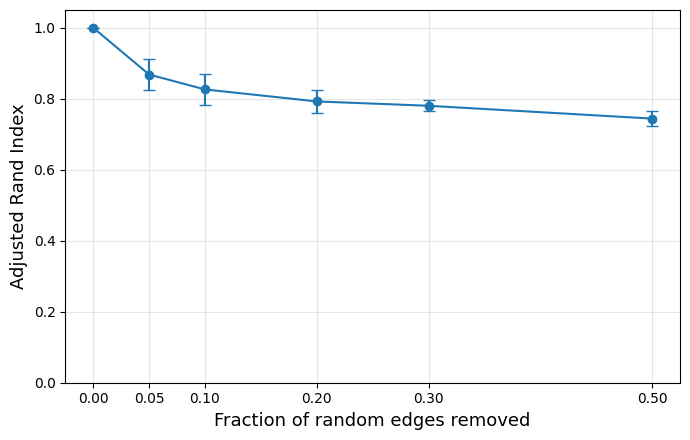}
    \caption{Agreement (ARI) between the perturbed and full-data clusterings at $M=6$ under random edge removal. Markers show the mean across 20 replicates; vertical bars show $\pm 1$ standard deviation.}
    \label{fig:app_missing}
\end{figure}

The clustering is very robust under random edge removal, ARI stays above 0.80 up to 10\% of edges removed, holds near 0.78 through 30\%, and remains around 0.74 even at 50\%. The overall partition structure is recovered reliably across all tested fractions, with only the boundaries between adjacent mid-tier clusters losing clarity at the highest removal rates. This stability confirms that the role assignments produced by the framework are dependable under the type of diffuse, non-targeted reporting imperfections most commonly encountered in supervisory practice.

The aggregate ARI conceals substantial heterogeneity in how individual roles withstand data loss, which Table~\ref{tab:app_f1} makes explicit. The most resilient roles are the green and blue clusters, the systemic cross-segment intermediary and the peripheral unsecured supplier respectively, both of which retain a mean $F_1$ above 0.94 even when half the edges are removed. The most fragile is the purple cluster, the mid-tier secured specialist reached largely through indirect links, whose $F_1$ falls to around 0.31 at $p=0.50$.

This ordering aligns with the counterparty-access dimension of Table~\ref{tab:cluster_roles} rather than with connectivity or segment specialism. The green and blue clusters are precisely the two whose role rests on direct counterparty access, despite sitting at opposite ends of the connectivity scale, the green cluster being systemic and the blue cluster peripheral. The clusters that lose definition first are those identified by indirect reach. The mechanism is transparent for random removal: a path of length $k$ is destroyed whenever any one of its $k$ constituent edges is removed, so a given length-$k$ path survives removal at rate $p$ with probability $(1-p)^k$, which decays geometrically in $k$. Higher-order path-count features therefore erode far faster than first-order ones, so a cluster whose identity rests on direct links retains its signature longest. The same mechanism distinguishes the two indirect secured specialists: the brown cluster, whose first-order links are marginally more pronounced than the purple cluster's, remains the better recovered of the two throughout.

\begin{table}[htbp]
    \centering
    \caption{Per-cluster mean $F_1$ score against the full-data baseline at $M=6$, averaged over 20 replicates under random edge removal. Columns follow the clusters of Table~\ref{tab:cluster_roles}. The case $p=0$ is omitted as it is identically one by construction.}
    \label{tab:app_f1}
    \begin{tabular}{l c c c c c c}
        \toprule
        $p$ & \textbf{Red} & \textbf{Green} & \textbf{Purple} & \textbf{Brown} & \textbf{Orange} & \textbf{Blue} \\
        \midrule
        0.05 & 0.902 & 0.975 & 0.726 & 0.670 & 0.735 & 0.998 \\
        0.10 & 0.798 & 0.974 & 0.522 & 0.586 & 0.737 & 0.994 \\
        0.20 & 0.728 & 0.966 & 0.329 & 0.592 & 0.675 & 0.991 \\
        0.30 & 0.645 & 0.960 & 0.332 & 0.604 & 0.693 & 0.991 \\
        0.50 & 0.467 & 0.948 & 0.306 & 0.598 & 0.589 & 0.990 \\
        \bottomrule
    \end{tabular}
\end{table}

\end{document}